\documentclass[aps,pra,reprint,amsmath,amssymb,superscriptaddress,nofootinbib,floatfix]{revtex4-2}
\usepackage{graphicx}
\usepackage{booktabs}
\usepackage{makecell}
\usepackage{xcolor}
\usepackage{microtype}
\usepackage[colorlinks=true,linkcolor=blue!50!black,citecolor=blue!50!black,urlcolor=blue!50!black]{hyperref}

\newcommand{\dev}[1]{\texttt{#1}}
\newcommand{\lG}{\lambda_G}
\newcommand{\lL}{\lambda_L}
\newcommand{\lM}{\lambda_M}
\hypersetup{pdftitle={Separability does not certify zero-noise extrapolation: factorial diagnostics in dynamic quantum circuits},pdfauthor={Aryan Kumar}}

\setcellgapes{1pt}

\begin{document}
\renewcommand{\dbltopfraction}{0.9}
\renewcommand{\dblfloatpagefraction}{0.8}
\setcounter{dbltopnumber}{2}
\raggedbottom

\title{Separability does not certify zero-noise extrapolation:\texorpdfstring{\\}{ }
Factorial diagnostics in dynamic quantum circuits}

\author{Aryan Kumar}
\affiliation{Amity International School, Sector-46, Gurugram, India}

\begin{abstract}
Separately amplifying gate, latency, and mid-circuit measurement errors can simplify zero-noise extrapolation, but a weak interaction between amplification knobs does not establish an accurate zero-noise limit. We distinguish observable separability, response-model adequacy, and calibration of that limit, and test them in a pre-specified $2^3$ factorial experiment on three IBM dynamic circuits. For the primary echoed spectator observable, the latency--measurement coefficient is $b_{LM}=+0.0007$ with a nominal 95\% bootstrap interval $[-0.0033,+0.0049]$. A separately planned follow-up meets a $\pm0.005$ equivalence criterion and recovers two injected interactions; both repeat runs nevertheless trigger the drift rule. Multi-knob extrapolations have smaller median absolute errors than the single-knob comparisons, yet all five original strategies exclude the known ideal value for three of six observables. Duration controls support a model-dependent contribution to the spectator deficit from unequal decay rates in native padding and added delay. We give an explicit construction showing why measurements at two measurement-knob levels cannot identify the extrapolated limit without response assumptions. A post-hoc precision analysis and an additive full-factorial estimator expose a further bias--variance trade-off hidden by point-error comparisons. The experiment provides a diagnostic of the implemented response surface, not a certification of physical channel independence or mitigation accuracy.
\end{abstract}

\maketitle

\section{Introduction}

Zero-noise extrapolation (ZNE) estimates an ideal quantum observable from deliberately noisier circuit executions~\cite{Temme2017,Kandala2019}. Its validity depends on the response to noise amplification and on whether the extrapolated coordinate corresponds to the intended noise-free circuit. Noise can be increased by stretching control pulses~\cite{Kandala2019} or, more conveniently, by ``folding'' gates: a gate $U$ is replaced by $UU^\dagger U$, which ideally does the same thing but roughly triples the gate noise~\cite{GiurgicaTiron2020,LaRose2022}.

When a circuit contains several kinds of noise, a natural generalization is to amplify each kind separately and fit a function of several noise parameters~\cite{Otten2019,Russo2024}. Whether the extra work pays off is debated: a published Comment argues that conventional extrapolation with a single noise parameter already works when the noise is not identically distributed~\cite{Etxezarreta2024}. Time-correlated noise adds a further complication: it is hard to amplify such noise without also changing its spectrum, especially when individual gates are amplified, although global folding can remain reliable~\cite{Schultz2022}.

Dynamic circuits raise this question in a sharp form. In a dynamic circuit some qubits are measured partway through, and later operations are chosen according to the outcomes, which classical hardware computes in real time~\cite{CarreraVazquez2024,Baumer2024a}. Such circuits suffer, among other errors, from three physically distinct kinds of noise: errors in the gates, decoherence of every qubit while it waits for the classical decision (the \emph{feedforward latency}), and errors caused by the mid-circuit measurements (MCMs) themselves. We will call these $G$, $L$ and $M$. Folding the gates reaches only the first.

Several groups have begun to amplify the other two. The idea of adding idle time, in the form of identity or delay pulses, to increase the noise before extrapolating goes back at least to Tomkins and de Sousa~\cite{Tomkins2020}. Carrera Vazquez \textit{et al.} stretched the latency of conditional operations with delays inside a staggered dynamical-decoupling sequence and extrapolated linearly to zero delay; without the decoupling, they observed oscillations that prevented an accurate extrapolation~\cite{CarreraVazquez2024}. Santos and Uzdin amplified measurement errors by repeating each measurement and acting on the parity of the outcomes~\cite{Santos2025}. Bar, Santos and Uzdin extended layered gate mitigation to dynamic circuits while leaving the mid-circuit measurements unscaled~\cite{Bar2026}, and Uzdin has proposed a virtual noise-scaling method described as compatible with dynamic circuits~\cite{Uzdin2026}. Shirgure and Niu pointed out that globally folding the dynamic parts of a circuit scales all three kinds of noise together, with a single scale factor~\cite{Shirgure2026}. And Farrell \textit{et al.}, running a 139-qubit dissipative protocol with mid-circuit resets on \dev{ibm\_boston}, used ZNE that amplified only two-qubit gates and idling, and found that it did not effectively mitigate the increased noise sensitivity of their circuits with fewer environment qubits; they identify imperfect mid-circuit resets as a secondary source of noise~\cite{Farrell2026}.

Other work characterizes or cancels dynamic-circuit noise instead of extrapolating it, through probabilistic error cancellation~\cite{Gupta2024}, readout mitigation for mid-circuit measurements~\cite{Koh2026}, learning and characterization of the noise of measurements and feedforward~\cite{Hines2025,Chu2026,Edwards2026}, error-aware compilation~\cite{Zhong2025} and learned dynamical decoupling~\cite{Tong2026}. In a study of decoupling, Rahman, Egger and Arenz used repeated mid-circuit measurements to amplify measurement-induced noise on neighbouring qubits~\cite{Rahman2024}.

Suppose, then, that we can amplify $G$, $L$ and $M$ separately. Must we amplify them jointly and fit a model with many parameters, or may we amplify one at a time and simply combine the results? The answer depends on whether the three kinds of noise are \emph{separable}: whether each one reduces an observable by a factor that does not depend on the others. If they are, the effects multiply, the logarithm of the observable is a sum of separate terms, and separate one-dimensional extrapolations can be combined, provided each extrapolation is itself valid. If they are not\,---\,if, for example, the tones used to measure one qubit also dephase an idle neighbour while it waits for the feedforward\,---\,then the effects do not multiply, and omitting the resulting interactions can bias the combined extrapolation. Separability in this sense is a property of the observable and the knobs, not of the physical channels: channels that act quite independently need not combine multiplicatively, as Sec.~\ref{sec:coherent} will show. Our focus is a direct factorial diagnostic of this assumption for separately controlled gate, latency, and measurement operations in small dynamic circuits; we do not claim that multivariate extrapolation or factorial interaction analysis is new.

We separate three questions that are often conflated: whether the observable is separable over the settings measured, whether the extrapolation model remains valid outside those settings, and whether its zero coordinate removes the relevant noise. We do not propose a new way of amplifying noise; our three amplification ``knobs'' are adapted from the work cited above. Instead we ask, with an experiment designed for the purpose, whether the noise they amplify is separable on real hardware. The experiment is a factorial design: each knob is set low or high, in all eight combinations, on three small dynamic circuits whose ideal outputs are known exactly, and the analysis was fixed in advance. Along the way we derive simple formulas for the bias of one-at-a-time and of uniform extrapolation in terms of the interactions, and we point out that coherent phase errors can masquerade as interactions\,---\,a pitfall that our own first trial run illustrates.

In brief, we find that once the idle periods are echoed, the interaction we considered most likely is consistent with zero at the resolution of the experiment; two later runs, under different calibrations, also found it small, and the analysis recovered interactions of known size that we injected on the hardware. Simple one-at-a-time or uniform extrapolation does about as well as a full joint fit, and in two of our three circuits, amplifying every kind of noise matters far more than how the amplifications are combined. But separability is not accuracy: for half of our observables, all five original extrapolation strategies exclude the ideal value. Duration controls support a contribution from the latency knob under an explicit slot-cost model. An additional post-hoc analysis gives a constructive identifiability limit, exact estimator weights, and an additive fit using the complete cube. These additions sharpen the interpretation of the existing experiment; they are not new confirmatory hardware tests.

We begin in Sec.~\ref{sec:theory} with a simple model of how the three kinds of noise might combine, and work out what goes wrong when they do not combine multiplicatively. Section~\ref{sec:methods} describes how each kind of noise is amplified, the circuits, and the plan of the experiment. Section~\ref{sec:results} reports what we found: first in trial runs that test the knobs, then in the main experiment, with a positive control, in a comparison of extrapolation strategies and in two repeat runs, and finally in two checks, made after the main runs, on our error bars and on the latency knob. Section~\ref{sec:discussion} discusses what the results do and do not show, and Sec.~\ref{sec:conclusion} contains concluding comments.

\section{Observable response, interactions, and identifiability}
\label{sec:theory}

\subsection{A simple model}

Consider first a circuit with only one kind of noise, amplified by a factor $\lambda$, where $\lambda=1$ is the native circuit. If each unit of noise multiplies the expectation value $E$ of an observable by the same factor, then $E(\lambda)=E(0)\,e^{-\gamma\lambda}$, and the extrapolated value $E(0)$ follows from a straight-line fit of $\ln E$ against $\lambda$. A sufficient condition is that the measured observable is an eigenoperator of the adjoint noise channel, with a positive eigenvalue, and amplification repeats that channel without changing the ideal operation. A single Markovian decay mode also has this form. An arbitrary observable under a Pauli channel, or a sum of decay modes, need not follow one exponential.

Now let there be three kinds of noise, amplified by $\lG$, $\lL$ and $\lM$. The simplest possibility is that each reduces the observable by a factor of its own, whatever the other knobs are doing. Then $f=\ln E$ is additive: $f=f_0+\sum_i h_i(\lambda_i)$, with $h_i(0)=0$. This is observable separability. The stronger log-linear model takes $h_i(\lambda_i)=-\gamma_i\lambda_i$. We use that stronger model for two-point extrapolation; separability alone does not rule out curvature. The logarithmic formulation requires a positive observable throughout the fitted region. The analysis uses $E>0.05$ as a numerical guard and discards bootstrap values that fail it; this does not extend the model to zero crossings or negative expectations. It is a statement about how the knobs act on an observable, not about the physical channels: channels that act independently need not combine this way in general, although multiplicative attenuation holds in suitable Pauli-eigenoperator settings. Independent physical operations alone do not guarantee it, particularly after averaging over measurement branches. Here $f_0=\ln E_*$, where $E_*$ is the value the knobs extrapolate to at $\boldsymbol\lambda=\mathbf 0$. We write $E_*$ rather than $E_{\rm id}$, the ideal value, because the knobs reach only the noise they amplify; single-qubit gates, state preparation, terminal readout and, in our circuits, echo pulses are amplified by no knob, so in general $E_*\neq E_{\rm id}$. Under the log-linear separable model, each rate $\gamma$ can be measured by turning one knob at a time. But nothing guarantees separability. Suppose, for instance, that measuring one qubit also dephases its idle neighbour~\cite{Rahman2024}, and that this extra dephasing acts only during the feedforward window. Then the rate of decay during the window grows with $\lM$, and the total damage grows with the product $\lL\lM$. To allow for such effects we add products of the knobs to the exponent:
\begin{equation}
f(\boldsymbol\lambda)=f_0-\sum_i\gamma_i\lambda_i-\sum_{i<j}\gamma_{ij}\lambda_i\lambda_j-\gamma_{GLM}\,\lG\lL\lM ,
\label{eq:model}
\end{equation}
where $i,j\in\{G,L,M\}$. The coefficients $\gamma_{ij}$ and $\gamma_{GLM}$ measure interactions, and the noise is separable if they all vanish.

In a factorial experiment each knob takes two values, which we choose to be 1 and 3. It is then convenient to centre the knobs, $x_i=\lambda_i-2=\pm1$, and to write
\begin{equation}
f=b_0+\sum_i b_i x_i+\sum_{i<j}b_{ij}x_ix_j+b_{GLM}\,x_Gx_Lx_M .
\label{eq:coded}
\end{equation}
The eight coefficients $b$ are simply contrasts between the eight corners of a cube. The coefficient $b_{LM}$, for example, is one quarter of the change in the effect of latency when the measurement knob is turned up: if latency and measurement noise are separable, stretching the window costs the same, on the logarithmic scale, whether or not the measurements are repeated, and $b_{LM}=0$. Substituting $\lambda_i=2+x_i$ into Eq.~\eqref{eq:model} gives $b_{GLM}=-\gamma_{GLM}$ and $b_{ij}=-\gamma_{ij}-2\gamma_{GLM}$, so the interactions in Eq.~\eqref{eq:model} vanish exactly when those in Eq.~\eqref{eq:coded} do. Note, however, that with two levels per knob the eight coefficients fit the eight corners exactly, whatever the data. They measure non-additivity between the levels actually run, 1 and 3; that the same description holds down to $\boldsymbol\lambda=\mathbf 0$, where we extrapolate, is an assumption, which the centre points of Sec.~\ref{sec:curvature} test only in part.

\subsection{What goes wrong when the noise is not separable}

How much does an interaction matter? Consider the simplest component-resolved strategy, which we call \emph{one-at-a-time} (OAT) extrapolation. Starting from the native circuit, it turns up each knob alone to $\lambda_i=3$, reads off the slope $\hat\gamma_i=-[f(\mathbf e_i)-f(1,1,1)]/2$, where $\mathbf e_i$ denotes the setting with knob $i$ at 3 and the other two at 1, and estimates $\hat f_0=f(1,1,1)+\sum_i\hat\gamma_i$. For a log-linear separable response this is exact at the population level; exponentiating a noisy log estimate introduces additional finite-sample bias. Otherwise each measured slope also picks up the interactions that involve its knob, $\hat\gamma_i=\gamma_i+\sum_{j\neq i}\gamma_{ij}+\gamma_{GLM}$, and the estimate is off by
\begin{align}
\hat f_0^{\rm OAT}-f_0&=\sum_{i<j}\gamma_{ij}+2\gamma_{GLM}\nonumber\\
&=-(b_{GL}+b_{GM}+b_{LM})+4b_{GLM}.
\label{eq:oat}
\end{align}
A second strategy, \emph{uniform scaling}, turns all three knobs up together, $\lG=\lL=\lM=\lambda$, as global folding of dynamic circuits does~\cite{Shirgure2026}, and extrapolates exponentially from $\lambda=1$ and 3. Its error is
\begin{align}
\hat f_0^{\rm unif}-f_0&=3\sum_{i<j}\gamma_{ij}+12\gamma_{GLM}\nonumber\\
&=-3(b_{GL}+b_{GM}+b_{LM})+6b_{GLM}.
\label{eq:unif}
\end{align}
Within Eq.~\eqref{eq:model}, both strategies are exact at the population level when the interactions vanish, and Eqs.~\eqref{eq:oat} and~\eqref{eq:unif} tell us exactly how they fail when they do not; uniform scaling is three times as sensitive to pairwise interactions. A third strategy, \emph{joint} component-resolved ZNE (joint CR-ZNE), fits all eight coefficients of Eq.~\eqref{eq:coded} to the eight corners and evaluates Eq.~\eqref{eq:model} at $\boldsymbol\lambda=\mathbf 0$; it is exact for any interactions of the form~\eqref{eq:model}, at the price of measuring all eight corners. For comparison we will also try two strategies that amplify only one kind of noise, \emph{latency-only} and \emph{gate-only} extrapolation, which fit a straight line to $E$ over several levels of a single knob, the first as in Ref.~\cite{CarreraVazquez2024}. These leave other noise sources unscaled and therefore do not, in general, recover the ideal value. Their observed error also includes response-model and fitting effects; it cannot be attributed solely to the unscaled noise. We will call the first three, which amplify every kind of noise, the \emph{multi-knob} strategies. The derivations are in Appendix~\ref{app:derivations}.

One caution is in order. Because Eqs.~\eqref{eq:oat} and~\eqref{eq:unif} are algebraic identities on the same eight corners, comparing two strategies on one data set merely reproduces them. To test whether interactions predict bias, we will estimate the interactions from one half of the data and the biases from the other.

\subsection{What a factorial contrast identifies}
\label{sec:identifiability}
At the eight corners, the contrast coefficients in Eq.~\eqref{eq:coded} are defined without assuming a physical noise model. All four mixed coefficients must vanish for additivity on that cube. A small $b_{LM}$ alone is weaker. At a fixed gate setting $x_G$, define
\begin{equation}
R_{LM}(x_G)=\frac{E(x_G,+1,+1)E(x_G,-1,-1)}{E(x_G,+1,-1)E(x_G,-1,+1)}.
\end{equation}
The exact corner identity is
\begin{equation}
\tfrac14\ln R_{LM}(x_G)=b_{LM}+x_Gb_{GLM}.
\label{eq:conditional}
\end{equation}
Thus $b_{LM}$ is the gate-averaged log interaction. Opposite conditional interactions can cancel in it. Its $\pm0.005$ margin corresponds to a geometric-mean ratio $\sqrt{R_{LM}(-1)R_{LM}(+1)}$ between $e^{-0.02}$ and $e^{0.02}$, approximately 0.980 and 1.020; it does not impose that margin on either ratio separately.

Even perfect corner additivity does not identify the target. Let $f(\boldsymbol\lambda)$ be any response compatible with measurements made only at $\lambda_M\in\{1,3\}$. For any constant $c$, the response
\begin{equation}
f_c(\boldsymbol\lambda)=f(\boldsymbol\lambda)+c(\lambda_M-1)(\lambda_M-3)
\label{eq:nullspace}
\end{equation}
agrees at every sampled setting, including the latency scans and mixed gate centre points, but satisfies $f_c(\mathbf0)=f(\mathbf0)+3c$. The added term is univariate, so it changes no mixed interaction anywhere. This is a response-level identifiability limitation, not a claim that every continuation is generated by the same microscopic device model. Physical range constraints alone do not remove it: $f=-0.05(\lambda_G+\lambda_L+\lambda_M)$ and its continuation with $c=-0.02$ both give $0<E\le1$ for nonnegative knobs, agree at both measurement levels, and have zero-coordinate values 1 and $e^{-0.06}\simeq0.942$. No amount of additional shots at these levels separates them. These illustrative responses are not hardware fits.

The construction is an elementary consequence of finite-node extrapolation, the interpolation setting also underlying layerwise Richardson methods~\cite{Russo2024}. Its role here is to distinguish an experimentally estimable interaction from an assumed continuation to zero. Additional physical measurement levels can test the quadratic ambiguity; a finite grid still requires a response class or regularity assumption to determine the unsampled limit. Calibration of what the zero coordinate means remains a separate requirement.

\subsection{Extrapolation weights and statistical cost}
\label{sec:precision}
Write a log estimate as $\hat f_0=\sum_{\mathbf s}w_{\mathbf s}\hat f(\mathbf s)$. Joint extrapolation is tensor-product interpolation with one-axis weights $w(1)=3/2$ and $w(3)=-1/2$, so $w_{\mathbf s}=\prod_iw(s_i)$. OAT assigns weight $5/2$ to $(1,1,1)$ and $-1/2$ to each of its three single-knob neighbours. Uniform extrapolation assigns $3/2$ and $-1/2$ to the two diagonal corners.

For a comparison that also uses all eight corners but imposes additivity, we introduce an unweighted additive log fit,
\begin{equation}
\hat f_0^{\rm add}=b_0-2(b_G+b_L+b_M),\qquad
w_{\mathbf s}^{\rm add}=\frac{1-2\sum_i x_i}{8}.
\label{eq:additive}
\end{equation}
This standard factorial main-effects projection was not in the original analysis plan. Within the multilinear model its error is
\begin{equation}
\hat f_0^{\rm add}-f_0=-4\sum_{i<j}b_{ij}+8b_{GLM}.
\label{eq:additivebias}
\end{equation}
Reducing variance by omitting interactions therefore introduces an explicit bias, rather than making interactions irrelevant.

For independent settings with per-shot log-variance factor $v_{\mathbf s}$, the delta method gives
\begin{equation}
\operatorname{Var}(\hat f_0)\simeq\sum_{\mathbf s}\frac{w_{\mathbf s}^2v_{\mathbf s}}{N_{\mathbf s}}.
\label{eq:variance}
\end{equation}
For independent $\pm1$ shots, $v_{\mathbf s}=(1-E_{\mathbf s}^2)/E_{\mathbf s}^2$; clustered twirls or drift require a corresponding covariance model. At fixed total shots $N_{\rm tot}$, allocation $N_{\mathbf s}\propto|w_{\mathbf s}|\sqrt{v_{\mathbf s}}$ minimizes this variance and gives $(\sum_{\mathbf s}|w_{\mathbf s}|\sqrt{v_{\mathbf s}})^2/N_{\rm tot}$. These are shot-budget statements, not equal-QPU-time comparisons: circuit durations differ.

Table~\ref{tab:variance} isolates the design cost under equal $v_{\mathbf s}=v$. Equal shots per required setting give factors 125, 28, 5, and 13 multiplying $v/N_{\rm tot}$ for joint, OAT, uniform, and additive fits, respectively. The 25-fold joint-to-uniform variance ratio is conditional on this homogeneous model, not an empirical performance ratio. Heterogeneous responses and the risk of model bias matter in practice.
\begin{table}[t]
\caption{Analytic log-variance factors under independent settings and equal per-shot factor $v$. $K$ is the number of required settings. $F_{\rm eq}=K\sum w^2$ and $F_{\rm opt}=(\sum|w|)^2$ multiply $v/N_{\rm tot}$. This post-hoc comparison excludes setting-dependent runtime and drift.}
\label{tab:variance}
\centering
\begin{tabular}{lrrr}
\toprule
Estimator & $K$ & $F_{\rm eq}$ & $F_{\rm opt}$\\
\midrule
Joint & 8 & 125 & 64\\
OAT & 4 & 28 & 16\\
Uniform & 2 & 5 & 4\\
Additive cube & 8 & 13 & 9\\
\bottomrule
\end{tabular}
\end{table}

\subsection{When nothing interacts but something seems to}
\label{sec:coherent}

There is a subtlety that will turn out to matter. Suppose an idle qubit accumulates a coherent phase $\varphi$, so that $E\propto\cos\varphi$, and suppose each knob adds to the phase independently, $\varphi=\varphi_0+\sum_i a_i x_i$. Nothing here interacts; the phases simply add. Yet $\ln\cos\varphi=-\varphi^2/2-\varphi^4/12-\dots$ contains the cross term $-a_La_Mx_Lx_M$, so to leading order
\begin{equation}
b_{LM}=-a_La_M ,
\label{eq:coh}
\end{equation}
and likewise for every pair of knobs that both shift the phase. On the logarithmic scale, independent coherent errors produce interactions, and an apparent interaction is therefore not, by itself, evidence that the noise channels interact physically. An ideal echo that fully refocuses a static latency-dependent phase sets $a_L\to0$ and removes this contribution. Real echoes can leave residual phase, pulse errors, and time-dependent coupling; suppression of a contrast is not a unique diagnosis of its physical source.

\subsection{Curvature}
\label{sec:curvature}

Finally, the knobs need not act linearly even when nothing interacts; non-Markovian dephasing, for instance, makes noise grow non-linearly with the length of a delay~\cite{Schultz2022}. If a knob also enters quadratically, $f\supset k_i\lambda_i^2$, then each multi-knob strategy, which sees only the levels 1 and 3, extrapolates the quadratic term as if it were linear, and misses by $-3k_i$. A centre point, with $\lG=\lL=2$, reveals such curvature. The curvature $c$, the difference between $f$ at the centre and the mean of $f$ over the four corners with the same $\lM$, equals $-(k_G+k_L)$, because all interaction terms cancel in this contrast. The three multi-knob estimates are therefore shifted by
\begin{equation}
\Delta\hat f_0=3c-3k_M ,
\label{eq:curv}
\end{equation}
where $k_M$ cannot be determined with only two levels of $\lM$ (Appendix~\ref{app:derivations}).

\begin{figure*}[t]
\centering
\begin{minipage}[b]{0.60\textwidth}\centering
\includegraphics[width=\linewidth]{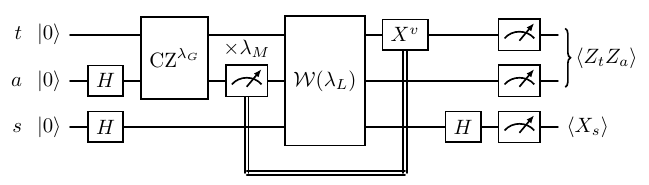}\\[-2pt]
{\small (a)}
\end{minipage}\hfill
\begin{minipage}[b]{0.35\textwidth}\centering
\includegraphics[width=\linewidth]{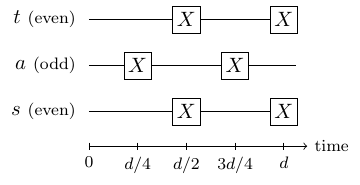}\\[-2pt]
{\small (b)}
\end{minipage}
\caption{(a) The spectator probe, our primary circuit. A CZ gate, folded $\lG$ times, exposes gate noise. The ancilla $a$ is measured $\lM$ times, and the parity $v$ of the outcomes conditions an X correction on $t$. The box $\mathcal W(\lL)$ stands for the rest of the feedforward window, during which all three qubits wait; the nominal window coordinate is $\lL\tau_w$, excluding fixed pulse overhead and any unmeasured scheduling latency. The spectator $s$ takes no part in the computation; $\langle X_s\rangle$ records how much it dephases. (b) The staggered echo inside the window: qubits at even and odd positions along the chain receive their X pulses at different times, which refocuses static Z rotations and static ZZ coupling.}
\label{fig:circuit}
\end{figure*}

\section{The experiment}
\label{sec:methods}

\subsection{Turning up one kind of noise at a time}
\label{sec:knobs}

How can each kind of noise be amplified without disturbing the others? Our knobs manage this only approximately, and throughout, $\lambda$ denotes a knob's nominal setting: how many times a CZ is applied, how far the window is stretched, how many times a measurement is repeated. That the noise grows in proportion to it is an assumption, which Sec.~\ref{sec:pilots} shows to fail for the measurement knob. For the gates, the standard method is folding. We replace each CZ gate by $\mathrm{CZ}^{\lG}$, an odd number of repetitions that ideally acts as a single CZ but multiplies the gate noise roughly $\lG$-fold, with every CZ randomly Pauli-twirled and with barriers that stop the compiler from cancelling the repetitions. Folding also lengthens the time before the measurement, during which the other qubits wait, so the gate knob amplifies some idle noise as well; we did not compensate for this.

For the latency, we lengthen the window during which the qubits wait for the feedforward. Following Carrera Vazquez \textit{et al.}~\cite{CarreraVazquez2024}, who stretch a duration $\tau$ by a factor $c=(\tau+\delta)/\tau$ and extrapolate to $c=0$, we add free evolution of length $(\lL-1)\tau_w$, so that its nominal duration coordinate is $\lL\tau_w$. Identifying $\lL=0$ with a physical zero-length window requires the assumed baseline timing and scaling model. Here $\tau_w=3t_{\rm slot}+\tau_{\rm ff}$, where $t_{\rm slot}=t_{\rm meas}+2t_X$ is the length of one measurement slot and $\tau_{\rm ff}$ is the feedforward latency. Because $\tau_{\rm ff}$ is not published for each device, we assume $\tau_{\rm ff}=1.0~\mu$s. This choice changes what $\lL=0$ means in physical time, but not the interaction estimates, which compare only the two levels that were actually run; Sec.~\ref{sec:extrap} shows how much it matters for the extrapolations. More generally, the interaction estimates rest only on the levels run, whereas every extrapolation to $\lL=0$ also assumes that the window as run decoheres at the same rate as the free evolution added to stretch it (Sec.~\ref{sec:discussion}).

A first trial run, described in Sec.~\ref{sec:pilots}, showed that an unprotected window lets an idle qubit rotate coherently, which by Eq.~\eqref{eq:coh} is exactly what we must avoid. Every window therefore carries a fixed \emph{staggered echo} [Fig.~\ref{fig:circuit}(b)], similar in spirit to the staggered decoupling sequence of Ref.~\cite{CarreraVazquez2024}: qubits at even positions along the chain receive X pulses at $d/2$ and $d$, and those at odd positions at $d/4$ and $3d/4$, where $d$ is the total free evolution. For ideal pulses, this refocuses static Z rotations on every qubit and static ZZ coupling between neighbours. Each qubit receives two pulses in each explicit delay sequence at every setting, so their count does not scale with $\lL$. The knob changes the programmed free-evolution duration; finite pulse widths and residual coherent or time-dependent errors need not obey a pure dephasing-and-relaxation model.

For the measurements, we follow Santos and Uzdin~\cite{Santos2025}: each mid-circuit measurement that feeds a decision is repeated $\lM$ times, and the correction is conditioned on the parity of the outcomes. Here there is a trap. Repeating a measurement takes time, so turning up $\lM$ would also lengthen the window, and measurement noise would secretly carry latency noise with it. To prevent this, we pad every $\lM=1$ window with $2t_{\rm slot}$ of echoed idling. This matches the nominal slot-duration budget at each latency setting, making the measurement knob compare measurement operations with padded waiting. It is not a measurement of the backend's realized schedule. Conditional control latency, twirl-dependent pulses, and compiler scheduling can leave timing differences, especially in circuits with several measurements and conditions. The records validate inserted delays and instruction counts, not the full executed pulse timeline; all separability claims therefore concern the implemented knobs.

All twirling was applied by hand and recorded, because IBM's built-in measurement twirling does not cover measurements that feed conditions. Mid-circuit measurements use the dedicated instruction \texttt{measure\_2}, and terminal measurements the standard \texttt{measure}.

\subsection{Three circuits with known answers}

Mitigation can be judged only against a known answer, so we chose three small dynamic circuits whose ideal expectation values are all exactly $+1$. The primary one, the \emph{spectator probe} of Fig.~\ref{fig:circuit}(a), is built to expose the interaction we considered most likely. Three qubits sit in a row: a target $t$, an ancilla $a$ and a spectator $s$, with $a$ and $s$ prepared in $|{+}\rangle$. A CZ gate between $a$ and $t$, ideally the identity because $t$ starts in $|0\rangle$, exposes gate noise. The ancilla is then measured mid-circuit, the window runs, and an X correction is applied to $t$ according to the outcome. The spectator's only job is to record how much it dephases while its neighbour is measured and the feedforward is computed. We measure $\langle X_s\rangle$, the primary observable, and $\langle Z_tZ_a\rangle$, which tests the feedforward itself.

The second circuit teleports a qubit twice: an $R_y(\pi/3)$ state on $q_0$ is passed to $q_2$ and then to $q_4$ through two Bell pairs, with two feeding measurements and conditional X and Z corrections per hop, and is then rotated back, so that ideally $\langle Z_4\rangle=1$; a spectator $q_5$ sits in $|{+}\rangle$, and we measure its $\langle X_5\rangle$. The third prepares a four-qubit GHZ state in constant depth, using measured parities and corrections conditioned on their running XOR~\cite{Baumer2024a}, with a spectator at the end of the chain; we measure the parity $\langle X^{\otimes4}\rangle$ and the spectator's $\langle X\rangle$. In all, there are six observables.

\subsection{Design}

Each knob takes the values 1 and 3, and all eight combinations, the corners of a cube, are run for each circuit. Two centre points, $(2,2,1)$ and $(2,2,3)$, test for curvature. Because $\lM$ must be odd (for even $\lM$ the parity of identical outcomes would always be zero), it has no centre level, so the centre point is repeated at both of its levels. The value $\lG=2$ is realized by folding all the CZs of a twirled instance of the circuit, or none of them, each with probability $1/2$. This is a mixture of endpoint circuits, not a physical intermediate gate-noise level. It is linear in $E$, not in $\ln E$, and therefore cannot identify gate-axis curvature. To first order in the gate contrast and under the quadratic response model, the centre contrast measures $c\simeq-k_L$, giving the approximate shift $3c-3k_G-3k_M$. Mixture imbalance and the log-of-mixture correction are quantified in Appendix~\ref{app:derivations}; a measured centre contrast should not be read as a model-free estimate of latency curvature. A comparison set adds the latency-only settings $(1,2,1)$, $(1,4,1)$ and $(1,5,1)$ and the gate-only setting $(2,1,1)$. Each setting of the main experiment used $2\times10^4$ shots, divided among 20 random twirls; the trial runs used $5\times10^3$ shots per setting, on the probe alone.

How small an interaction could this detect? An observable with values $\pm1$, estimated from $N$ shots, has a standard error of at most $1/\sqrt N$. A coded coefficient such as $b_{LM}$ is half the difference between the mean values on two halves of the cube, so at $N=2\times10^4$ its standard error is at most 0.0025, or $0.0025/E_{\min}$ on the logarithmic scale, where $E_{\min}$ is the smallest of the eight values. The trial run on the main device, scaled to the same number of shots, gave a standard error of 0.0016 for $b_{LM}$. From this we computed, before the main experiment, that an interaction of 0.0044 would be detected with 80\% probability in a two-sided test at the 5\% level (0.0052 at the most stringent Holm level); we call 0.0044 the minimum detectable effect.

\subsection{Pre-specified analysis}
\label{sec:plan}

The analysis was specified before collecting the confirmatory data; the trial runs informed the design. The plan named one primary hypothesis, H1: that $b_{LM}$ is non-zero for the probe spectator, judged by a bootstrap 95\% confidence interval together with a Holm correction over the four interaction terms. Three secondary hypotheses asked whether $b_{LM}$ exceeds the other pairwise interactions (H2), whether the extra bias of uniform over joint extrapolation grows with the size of the interactions across the six observables (H3, using independent halves of the data), and whether $b_{LM}$ is stable across days (H4). An exploratory comparison (E1) asked how much the echo reduces the apparent interaction. The plan also fixed the logarithmic scale as primary; bootstrap intervals with at least 2000 replicates, resampling twirls and shots; a drift rule that sets aside, for separate analysis, any later run in which the reported readout or CZ error of a qubit in use changes by more than a factor of two, its $T_1$ or $T_2$ falls by more than half, or the value at the native setting $(1,1,1)$ shifts by more than five standard errors; and an automatic trial-run validation on the trial runs.

\subsection{Hardware and follow-up design}
\label{sec:runs}

The experiment ran on IBM Quantum processors through Qiskit Runtime (Qiskit 2.5.2, qiskit-ibm-runtime 0.50.0). The first two trial runs used \dev{ibm\_kingston}. The main experiment used \dev{ibm\_marrakesh}, preceded by an echoed trial run with the same validation criteria except for the measurement-knob tests, whose circuits were not repeated. On \dev{ibm\_marrakesh} the circuits used physical qubits 4--11 in chain order: 4--6 for the probe, 4--9 for teleportation and 4--11 for the GHZ state. Each circuit was compiled onto this fixed chain, and an automatic check confirmed that compilation added no two-qubit gates and kept every measurement, conditional operation and delay. After analysing these runs, we planned a follow-up on the probe with three aims: to repeat the primary test days later, to inject interactions of known size, and to measure the cost of the padding at two settings of the latency knob. In one batch, in random order, it ran the eight corners as in the main experiment; the same corners with two doses of an injected interaction (Sec.~\ref{sec:separability}), at $10^4$ shots per setting; and the settings $(1,1,1)$ and $(1,3,1)$ without padding, at $2\times10^4$. The follow-up analysis was specified before data collection, and the experiment used the same qubits more than four days after the main run. Calibration details are given in Appendix~\ref{app:runs}.

\begin{figure*}[t]
\includegraphics[width=0.92\textwidth]{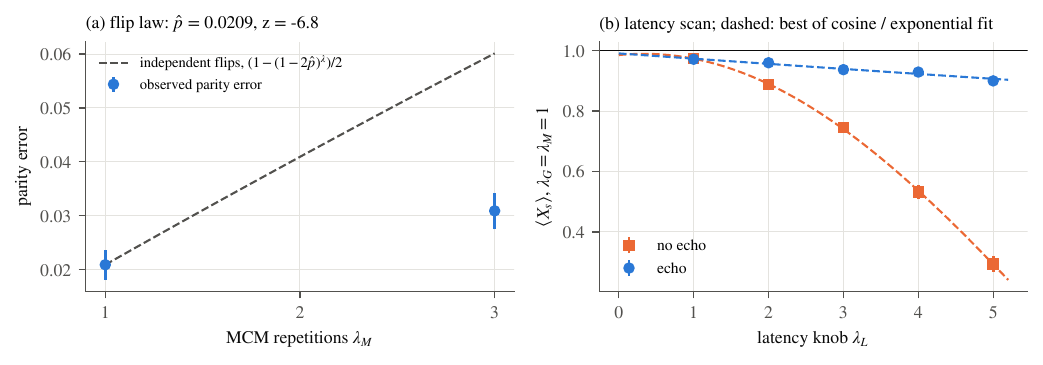}
\caption{What the trial runs on \dev{ibm\_kingston} revealed. (a) For an ancilla prepared in $|0\rangle$ or $|1\rangle$, the parity of $\lM$ repeated measurements is wrong far less often than independent errors would imply (dashed line), rejecting the identical independent bit-flip model . (b) Without an echo (orange), the spectator's $\langle X_s\rangle$ follows a cosine as the feedforward window is stretched, the signature of a coherent rotation; with the echo (blue) it decays slowly. Dashed lines show the fits drawn by our analysis: a cosine for the un-echoed trial and an exponential for the echoed trial (a slow cosine fits the echoed trial about as well).}
\label{fig:knobs}
\end{figure*}

\subsection{A check in simulation}

Before the hardware experiment, we ran the analysis on simulated data (Qiskit Aer) from a noise model with known coefficients; the numbers in this subsection are therefore simulated. With separable noise the analysis found $b_{LM}=-0.0008$ $[-0.0048,+0.0032]$, against a true value of zero, and raised no false alarm in 20 repetitions. With a synthetic latency\,$\times$\,measurement interaction added, it found $-0.0581$ $[-0.0633,-0.0533]$, against the true $-0.0562$, and detected the interaction in all 20 repetitions; the simulated estimators also reproduced Eqs.~\eqref{eq:oat} and~\eqref{eq:unif} to four decimal places. The simulation also shows how an interaction misleads uniform scaling: with the synthetic interaction, uniform extrapolation overshot the ideal value (1.0745, against 0.9014 for the joint fit), as Eq.~\eqref{eq:unif} predicts for a negative $b_{LM}$.

\section{What we found}
\label{sec:results}

The empirical results in this section are from IBM Quantum hardware; Sec.~\ref{sec:estimatorcomparison} presents a post-hoc reanalysis and an analytic variance comparison. Uncertainties are 95\% bootstrap confidence intervals; the hypothesis tests and the figures use separate bootstrap runs on the same data, so a few bounds differ in the last digit. For the follow-up, as its plan specified, they come from the instance-level bootstrap of Sec.~\ref{sec:bootstrap}.

\subsection{Do the knobs do what we intend?}
\label{sec:pilots}

The trial runs taught us two lessons. The first concerns the measurement knob. If each repetition of a measurement misreported the outcome independently with probability $p$, the parity of $\lM$ repetitions would be wrong with probability $[1-(1-2p)^{\lM}]/2$, nearly three times as often for $\lM=3$ as for a single measurement. To check this, the first trial run, on \dev{ibm\_kingston}, included circuits that prepare the ancilla in $|0\rangle$ or $|1\rangle$ and measure it once or three times. A single measurement erred with probability 0.0209, but the parity of three repetitions was wrong only 0.0309 of the time, far less than the 0.0601 that independence predicts, a difference of 6.8 standard errors [Fig.~\ref{fig:knobs}(a)]. Moreover, each successive repetition disagreed with the prepared state more often than the last (0.0192, 0.0253 and 0.0306). These observations reject the identical independent bit-flip model; correlated errors, repetition-dependent error rates, or measurement-induced state changes can contribute, consistent with the backaction studied in Ref.~\cite{Chu2026}. The measurement knob thus amplifies the effect of the measurement process as a whole, not simply readout error, and, as our plan specified, we interpret interactions involving $M$ accordingly.

The second lesson concerns the latency knob. The first trial run deliberately had no echo, so that stretching the window would add nothing but idle time, and it showed what Carrera Vazquez \textit{et al.} had warned of~\cite{CarreraVazquez2024}. As the window was stretched five-fold, the spectator's $\langle X_s\rangle$ fell from 0.97 to 0.29, but not exponentially, as Eq.~\eqref{eq:model} assumes; it followed a cosine almost perfectly [Fig.~\ref{fig:knobs}(b); $\chi^2=0.6$ for a cosine with 2 degrees of freedom, against 349.5 for an exponential with 3]. The spectator was rotating coherently, by about 0.27~rad per step of $\lL$, presumably because of a static frequency offset or ZZ coupling of the kind the echo is designed to cancel. A straight-line extrapolation of this curve to zero latency gives the impossible value 1.20. And, as Eq.~\eqref{eq:coh} warns, the run also showed an apparent interaction, $b_{LM}=+0.0198$ $[+0.0113,+0.0283]$, which, to leading order, Eq.~\eqref{eq:coh} alone would explain with $a_M\approx-0.07$~rad, given $a_L\approx0.27$~rad (we did not measure $a_M$ independently).

The echo suppressed both problems. On the same device and qubits, $\langle X_s\rangle$ now decayed slowly and smoothly, from 0.97 to 0.90 over the same range; an exponential fitted well ($\chi^2=4.0$ for 3 degrees of freedom; a cosine fits about as well, $\chi^2=2.5$ for 2, but only with a rotation five times slower than before, 0.054~rad per step); and the interaction became consistent with zero, $b_{LM}=-0.0014$ $[-0.0055,+0.0026]$. When the main experiment moved to \dev{ibm\_marrakesh}, a repeat of the echoed trial run there showed no rotation either ($b_{LM}=+0.0029$ $[-0.0032,+0.0090]$), although the decay flattened at the longest window and the exponential fit was less good ($\chi^2=19.6$ for 3 degrees of freedom). (These trial-run intervals come from the 500-replicate bootstrap of the trial-run validation.)

The trial runs also confirmed that duration matching was necessary. On \dev{ibm\_marrakesh}, padding a $\lM=1$ window with $2t_{\rm slot}$ of echoed idling lowered $\langle X_s\rangle$ by 0.064, ten times its standard error; without the padding, the measurement knob would have carried this much latency noise with it. (Section~\ref{sec:window} returns to this number.)

\subsection{Is the noise separable?}
\label{sec:separability}

\begin{figure*}[t]
\includegraphics[width=0.82\textwidth]{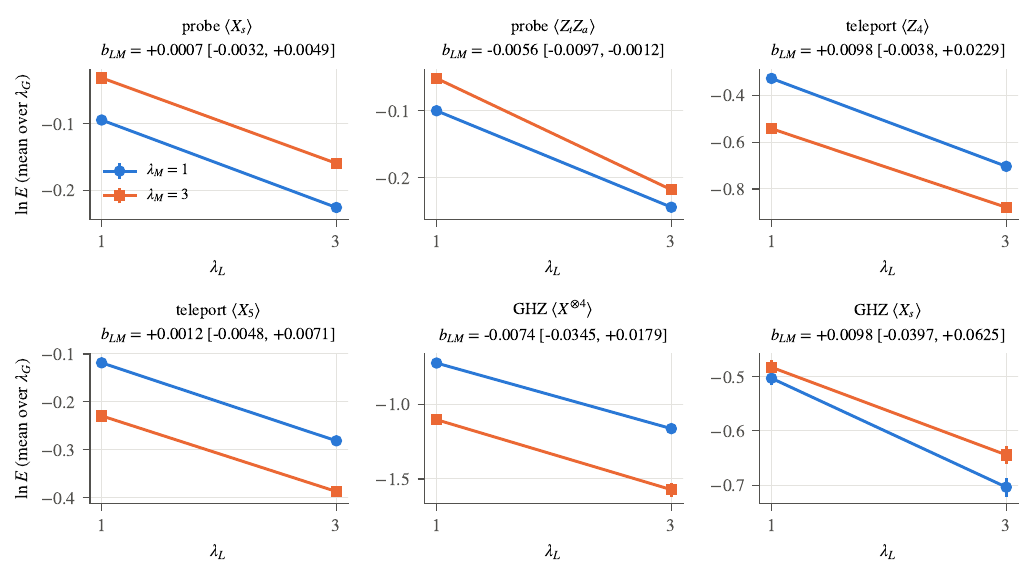}
\caption{The test of separability. Each panel shows $\ln E$ against the latency knob $\lL$, with the measurement knob at $\lM=1$ (blue) and $\lM=3$ (orange), averaged over $\lG$ (main experiment, \dev{ibm\_marrakesh}). If latency and measurement noise are separable, the two lines are parallel; their slopes differ by $2b_{LM}$, which is given with its 95\% interval above each panel.}
\label{fig:interaction}
\end{figure*}

With the knobs understood, the main experiment asked the central question. If latency and measurement noise are separable for the spectator, then on a logarithmic scale stretching the window must cost the same whether or not the measurement is repeated: the two lines in each panel of Fig.~\ref{fig:interaction} should be parallel within uncertainty. The probe spectator is compatible with this condition at the present resolution. The interaction is $b_{LM}=+0.0007$ $[-0.0033,+0.0049]$, and after the Holm correction none of the four interaction terms comes close to significance (adjusted $p$ values of 1, 0.8, 1 and 1 for $b_{GL}$, $b_{GM}$, $b_{LM}$ and $b_{GLM}$). The half-width of the interval, 0.0041, is smaller than the minimum detectable effect of 0.0044, which is the condition our plan set for declaring this interaction negligible. That rule is not a formal test of equivalence, however, and the interval does not exclude an interaction as large as $+0.0044$ itself. The nominal 95\% interval is contained in $(-0.005,+0.005)$ for this single, gate-averaged contrast. This is not a simultaneous bound on all interactions, a fixed-gate bound, or a bound on total extrapolation error. Under Eq.~\eqref{eq:model}, the contribution of this coefficient alone to the OAT log error is smaller than 0.005; the other coefficients, curvature, and target mismatch remain unconstrained by that statement. Nor is $b_{LM}$ larger than the other interactions, as H2 conjectured: $|b_{LM}|-|b_{GL}|=-0.0009$ $[-0.0045,+0.0036]$ and $|b_{LM}|-|b_{GM}|=-0.0019$ $[-0.0055,+0.0035]$.

Would the analysis have seen an interaction if there had been one? The minimum detectable effect answers this only on paper, so in the follow-up we injected interactions of known size. Just before the spectator is read out, we apply a virtual rotation about the $z$ axis, which IBM implements as a change of the reference frame for later pulses: by $\theta=a(x_L+x_M)$ in half of the instances and by $-\theta$ in the other half. For a common pre-rotation state averaged equally over the two signs, this multiplies $\langle X_s\rangle$ by exactly $\cos\theta$, whatever phase that state carries: it is the mechanism of Sec.~\ref{sec:coherent}, made exact in expectation. In finite samples, random twirls and temporal state variation can prevent exact cancellation of the sine term. It shifts $b_0$ and $b_{LM}$ by $\tfrac12\ln\cos2a$ and leaves the other six coefficients unchanged. We chose two doses, $a=0.0997$ and 0.1405~rad, which shift $b_{LM}$ by $-0.010$ and $-0.020$, two and four times the bound above. The measured shifts were $-0.0104$ $[-0.0148,-0.0058]$ and $-0.0213$ $[-0.0253,-0.0172]$, that is, 1.04 $[0.58,1.48]$ and 1.06 $[0.86,1.26]$ times the injected sizes, so the criterion fixed in the plan, that each interval exclude zero and contain the prediction, was met. At the corners where the rotation acted, $\ln\langle X_s\rangle$ fell by 0.022 and 0.016 for the smaller dose and by 0.038 and 0.040 for the larger, against 0.020 and 0.040 predicted, and none of the other six coefficients moved significantly, or by as much as 0.005. The analysis thus recovers interactions of this size, at the size injected.

One result of this run was unexpected. The feedforward observable $\langle Z_tZ_a\rangle$, which the rotation does not touch, should not have changed at all, yet its $b_{LM}$ was larger in the two rotated blocks than in the unrotated one, by $+0.0056$ $[-0.0010,+0.0124]$ and $+0.0080$ $[+0.0019,+0.0144]$. In a check made after the run, randomly exchanging instances between the blocks produced a difference as large as that of the two rotated blocks together, $+0.0068$, only 2\% of the time. The rotation can hardly be the cause. It changes only the reference frame of another qubit, after the last feedforward operation, and three quarters of the difference comes from the corners with $(\lL,\lM)=(3,1)$, where $\theta=0$, so that the rotated circuits there differ from the unrotated ones only in their random twirls; at those corners the spectator, too, differs between blocks, by $+0.0105$ $[-0.0002,+0.0204]$ for the larger dose. Nor is it a steady drift, although $\ln\langle Z_tZ_a\rangle$ did rise by about 0.02 over the two minutes of the run; removing that trend leaves the difference intact. The exploratory permutation $p$ value is conditional on exchangeability of the instances and is not the probability that the discrepancy arose by chance. Because several post-hoc diagnostics were examined, it is a warning about reproducibility rather than a separately confirmed physical effect. Differences of this size between independently built circuit sets call for caution.

The echo appears to have made the difference. Without it, in the first trial run, $|b_{LM}|$ was 0.0198; with it, in the main experiment, 0.0007, a reduction of 0.0191 $[0.0092,0.0272]$. Because these two runs also differ in device, calibration and number of shots, the cleaner comparison is between the two trial runs on the same device, in which $|b_{LM}|$ fell from 0.0198 to 0.0014.

\begin{figure*}[t]
\includegraphics[width=0.82\textwidth]{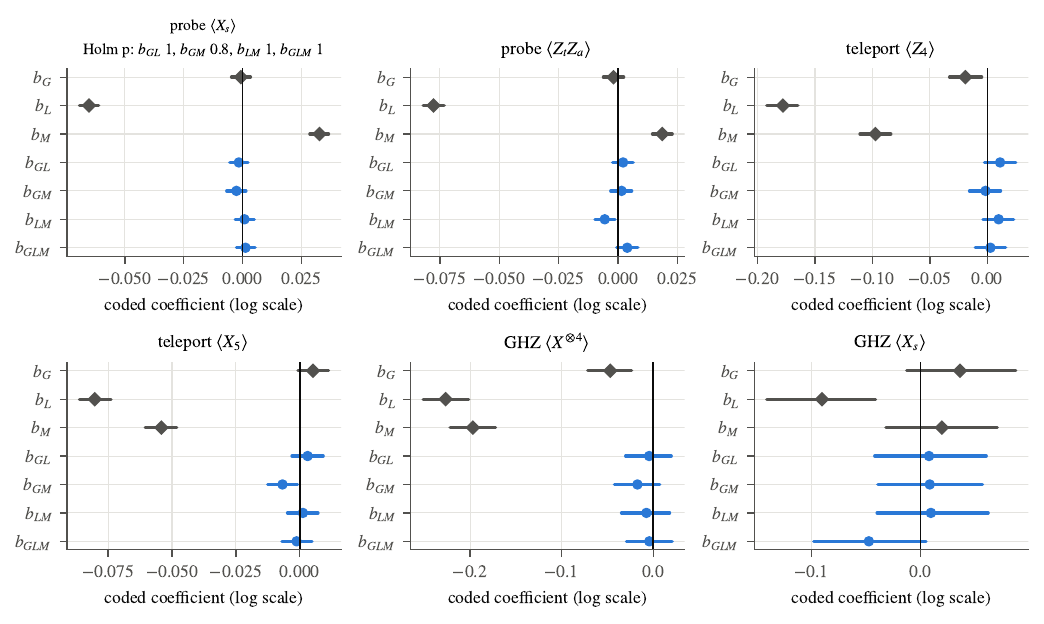}
\caption{Every coefficient of Eq.~\eqref{eq:coded} except $b_0$, with 95\% intervals, for the six observables (main experiment, \dev{ibm\_marrakesh}). Main effects are grey and interactions blue. Except for the poorly determined GHZ spectator, the interactions are small compared with the latency effect $b_L$. The probe $\langle X_s\rangle$ panel lists the Holm-adjusted $p$ values of the primary test.}
\label{fig:forest}
\end{figure*}

\begin{figure*}[t]
\includegraphics[width=0.84\textwidth]{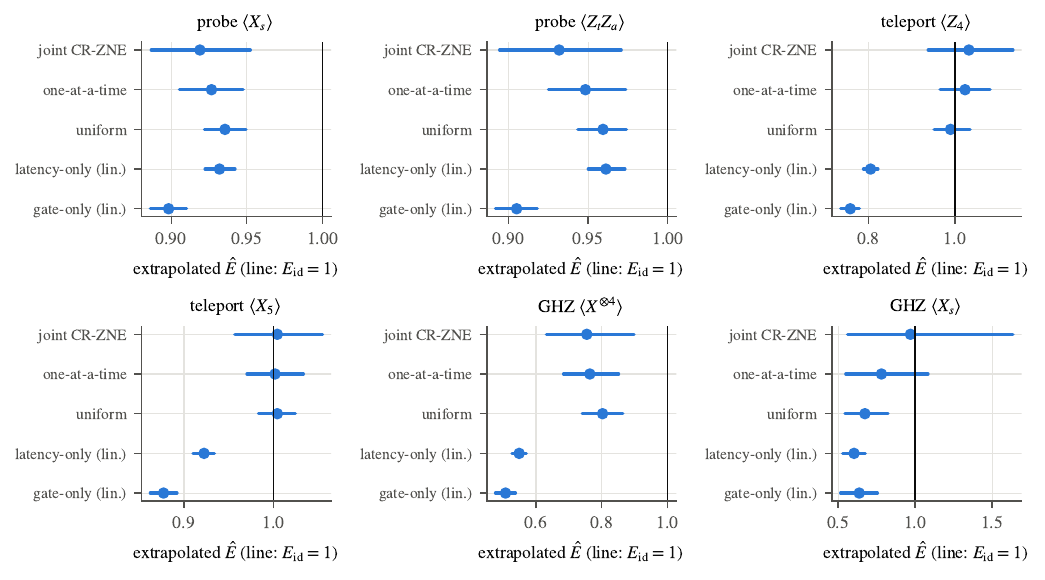}
\caption{Five ways to extrapolate to zero noise, compared with the exactly known ideal value $E_{\rm id}=1$ (vertical line) for each observable (main experiment and comparison set, \dev{ibm\_marrakesh}). The three multi-knob strategies (joint, one-at-a-time and uniform) broadly agree with one another. Latency-only and gate-only extrapolation fall well short for the teleportation and GHZ circuits; for the probe, latency-only extrapolation does about as well as the multi-knob strategies.}
\label{fig:est}
\end{figure*}

Figure~\ref{fig:forest} puts this result in context by showing every coefficient of Eq.~\eqref{eq:coded} but the constant $b_0$, for all six observables. The latency effect is the largest main effect for every observable, and it exceeds every interaction, by more than a factor of ten for all but the poorly determined GHZ spectator. Of the 24 interaction coefficients, two have intervals that exclude zero without correction: $b_{GM}$ for the teleportation spectator ($-0.0068$ $[-0.0126,-0.0009]$) and $b_{LM}$ for the probe's feedforward observable ($-0.0056$ $[-0.0097,-0.0012]$). About 1.2 such intervals would be expected by chance if all 24 coefficients were zero, and the two do not form a pattern; the second did not recur consistently in later runs (Sec.~\ref{sec:repeat}). Because the gate knob also adds idle time, interactions involving $G$ mix the effects of gates and of idling; the latency\,$\times$\,measurement interaction, our primary target, does not suffer from this. The GHZ intervals are too wide (up to about $\pm0.05$) to say much at the scale of the primary test.

One main effect deserves comment, because its sign is surprising. For the probe, turning up the measurement knob \emph{raises} both observables ($b_M=+0.0326$ for $\langle X_s\rangle$ and $+0.0186$ for $\langle Z_tZ_a\rangle$). Because of duration matching, however, the knob compares two extra measurements with an equal time of echoed idling, so a positive $b_M$ means only that, on this device, the idling did more harm than the measurements; the duration control had already shown that the padding alone lowers $\langle X_s\rangle$ by 0.064. For the teleportation observables and the GHZ parity, $b_M$ is negative, and for the GHZ spectator it is consistent with zero.

Finally, the centre points show that Eq.~\eqref{eq:model} is not exactly linear in $\lL$ for every observable. For the probe spectator the curvature excludes zero at both centre points, with nearly equal values ($-0.0144$ $[-0.0282,-0.0004]$ and $-0.0143$ $[-0.0261,-0.0030]$); for the teleportation spectator it excludes zero at one of them ($+0.0188$ $[+0.0055,+0.0329]$) and has the same sign at the other ($+0.0160$ $[-0.0023,+0.0348]$); and for the GHZ parity it only just excludes zero at $(2,2,1)$ ($-0.0575$ $[-0.1145,-0.0000]$), with the opposite sign at the other centre point. We return to the consequences in Secs.~\ref{sec:window} and~\ref{sec:discussion}.

\subsection{Which extrapolation should one use?}
\label{sec:extrap}

With the interactions known, we can ask the practical question: how well do the different strategies recover the ideal value? Figure~\ref{fig:est} shows the five estimates for each observable.

The original multi-knob strategies gave similar median point errors over these six observables. This descriptive comparison does not establish statistical equivalence or equal-budget performance. Their median absolute biases over the six observables were 0.0502 for the joint fit, 0.0625 for one-at-a-time and 0.0525 for uniform scaling. These errors are compatible with small interactions in several observables, but the medians conceal observable-specific differences and uncertainty. The two strategies that amplify only one kind of noise did worse, with median biases of 0.1366 (latency-only) and 0.1828 (gate-only), 2.6 to 3.6 times those of the joint and uniform estimates, and their intervals never contained the ideal value. (Our analysis also fitted both exponentially; latency-only extrapolation then has a median bias of 0.102, about twice that of the joint and uniform estimates, and gate-only extrapolation is almost unchanged.) For the teleportation and GHZ circuits, then, what matters most is not how the components are combined, but that all of them are amplified.

The probe is the exception. Its measured gate-knob effect is small, and here the multi-knob strategies gain nothing from the measurement knob: because turning that knob up raised the probe's observables, extrapolating it to zero lowers the estimate, by 0.036 on the logarithmic scale for the one-at-a-time $\langle X_s\rangle$, about half of that estimate's shortfall. For both probe observables, latency-only extrapolation did about as well as the multi-knob strategies.

Even the best strategies are not accurate everywhere, however. Their intervals include the ideal value for both teleportation observables, but for the probe spectator, the probe's feedforward observable and the GHZ parity, the 95\% intervals of every strategy lie below 1, and the multi-knob strategies miss by 0.04 to 0.25. (For the GHZ spectator, the joint and one-at-a-time intervals are too wide to judge.) With only six observables, the test of whether the extra bias of uniform scaling tracks the size of the interactions (H3) is uninformative: the Spearman correlation is $+0.14$, with interval $[-1,+1]$.

How much does the assumed feedforward latency matter? Our plan called for repeating the extrapolations with $\tau_{\rm ff}=0.5$ and 2~$\mu$s, and because the length of every window that was run is recorded, this needs no new data. On \dev{ibm\_marrakesh}, $t_{\rm slot}=1.956~\mu$s and $\tau_w=6.868~\mu$s, so a different latency moves the zero-window target from $\lL=0$ to $\lL=1-(3t_{\rm slot}+\tau_{\rm ff})/\tau_w$, that is, to $+0.07$ for 0.5~$\mu$s and to $-0.15$ for 2~$\mu$s. The joint and one-at-a-time estimates then move by at most 0.032, or 0.046 for the poorly determined GHZ spectator; the median biases of the multi-knob strategies stay between 0.050 and 0.067, against 0.126 to 0.142 for latency-only extrapolation; and no interval, computed with the pre-specified bootstrap, moves across the ideal value. Gate-only extrapolation does not involve $\lL$, and uniform scaling, which extrapolates all three knobs along a single line in nominal units, cannot be retargeted without separate information about the latency axis.

\subsection{Does the result hold up in repeat runs?}
\label{sec:repeat}

A repeat of the probe experiment on the same qubits, 2~h~37~min later, gave $b_{LM}=-0.0031$ $[-0.0056,-0.0004]$. Taken alone, this interval narrowly excludes zero, although it is a single uncorrected interval outside the family of the primary test. Its difference from the main run, $-0.0038$ $[-0.0088,+0.0011]$, is consistent with zero, which meets the plan's non-rejection rule for H4 (but see Sec.~\ref{sec:bootstrap}). Failure to reject a change is not evidence of stability without a stated equivalence margin. But the repeat run's baseline value had risen from 0.9058 to 0.9619, a shift of 15.7 standard errors, and the device's reported $T_1$ and $T_2$ had changed, which suggests a recalibration in between. The baseline shift alone triggers the drift rule, so, as planned, we report the repeat run separately rather than pooling it.

The follow-up measured the same eight corners again, with the same number of shots, more than four days after the main run. It gave $b_{LM}=-0.0014$ $[-0.0040,+0.0012]$, or $[-0.0050,+0.0023]$ with the bootstrap of our original plan, and its difference from the main run, $-0.0021$ $[-0.0061,+0.0020]$, is consistent with zero. For this run we had fixed a test of equivalence in advance: if its 90\% interval lay inside $\pm0.005$, the bound of the main run would be confirmed at the 5\% level. The interval, $[-0.0035,+0.0008]$, does lie inside, and so does the 90\% interval from the bootstrap of our original plan, $[-0.0044,+0.0017]$. This run, too, was flagged for drift: the ancilla's reported readout error had more than doubled since the main run, its reported error for mid-circuit measurements had grown from 0.0020 to 0.0186, and the baseline had shifted by 5.6 standard errors. As planned, we do not pool it with the others. In three runs under three calibrations, then, the primary interaction was $+0.0007$ $[-0.0024,+0.0038]$, $-0.0031$ $[-0.0049,-0.0013]$ and $-0.0014$ $[-0.0040,+0.0012]$, all three intervals from the instance-level bootstrap: small each time, though the first two differ by more than chance easily explains (Sec.~\ref{sec:bootstrap}).

The feedforward observable fared differently. Its $b_{LM}$, which excluded zero in the main run, was $+0.0017$ $[-0.0026,+0.0064]$ in the repeat run; in the follow-up it was $-0.0054$ $[-0.0096,-0.0013]$ in the unrotated circuits but $+0.0014$ $[-0.0021,+0.0049]$ in the rotated ones, which for this observable should be equivalent (Sec.~\ref{sec:separability}). The like-for-like repeat in the follow-up thus matched the main run, but the repeat run and the rotated circuits did not, and we do not regard that interaction as consistently reproduced.

\subsection{A check on the error bars}
\label{sec:bootstrap}

One choice in our plan deserves a second look. The bootstrap resamples the twirled instances of each circuit and then redraws the shots within each instance. But the spread between instances already contains the shot noise, so the second step counts it twice, and the intervals come out too wide: here by 20 to 40\%, except for the GHZ spectator, whose values vary between instances far more than shot noise alone would explain. As a check that was not part of the plan, we reran our analysis code with only that step removed, resampling whole instances, and with two different random seeds. Resampling whole instances is the usual bootstrap for data that come in clusters, and we regard it as the better estimate of the uncertainties; but because it was not in our plan, the pre-specified intervals remain primary for the trial runs, main experiment, comparison set, and first repeat throughout this paper, and we use the narrower intervals to strengthen none of our conclusions about them. (The plan for the follow-up, written afterwards, adopted the instance-level bootstrap; its main conclusion holds with either, Sec.~\ref{sec:repeat}.)

The primary result does not change: $b_{LM}=+0.0007$ $[-0.0022,+0.0037]$. Nor do the effect of the echo or the comparison of extrapolation strategies, where no interval moves across the ideal value. Three secondary statements do change. Four of the 24 interaction coefficients now exclude zero before correction, or five with one of the seeds: besides the two above, $b_{GLM}$ for the probe's feedforward observable ($+0.0039$ $[+0.0008,+0.0071]$), $b_{GL}$ for the teleportation observable ($+0.0111$ $[+0.0005,+0.0213]$) and, marginally, $b_{GLM}$ for the GHZ spectator; about 1.2 would be expected by chance. The difference between the repeat run and the main run, $-0.0038$ $[-0.0073,-0.0003]$, no longer contains zero, so H4 would fail by its own rule, in line with the drift flag on the repeat run. And the curvature now excludes zero at both centre points for the teleportation spectator, and clearly so at $(2,2,1)$ for the GHZ parity.

\subsection{Does the latency knob stretch the window faithfully?}
\label{sec:window}

Extrapolation to $\lL=0$ assumes that the window as it is run, padding and feedforward wait included, decays at the same rate as the free evolution that the latency knob adds to stretch it. If the window decays faster, the extrapolation removes too little of its noise, and the estimates come out low. Two kinds of data test this assumption for the probe: the shape of the latency axis, which we analysed after the runs, outside our plan, and the cost of the padding, which a trial run measured and the follow-up measured again, as planned.

The first is the shape of the latency axis. Together, the main experiment and comparison set ran every circuit at five latency levels, $\lL=1$ to 5, with $\lG=\lM=1$. For the probe spectator, $\ln\langle X_s\rangle$ fell by 0.073, 0.054, 0.039 and 0.034 in successive steps: the decay slows as the window grows, as the centre points had suggested. A quadratic through the five levels has the curvature $k_L=+0.0065$ $[+0.0028,+0.0098]$. Because the five levels come from two experimental runs, we use this coefficient, which, unlike a contrast between three of the levels, is not affected by a constant offset between the main experiment and comparison set. The echoed trial run on the same device, which ran all five levels in a single run, gives $+0.012$ $[+0.004,+0.019]$. By Eq.~\eqref{eq:curv}, curvature of this size shifts the multi-knob estimates by $-3k_L$, that is, by $-0.020$ or $-0.035$, against $-0.043$ (roughly $-0.085$ to $-0.001$) from the centre points; curvature along $\lG$ and $\lM$ would add to this. For the other observables the scans add little. They show no curvature for the probe's feedforward observable, the teleportation observable or the GHZ parity; for the GHZ spectator they agree in sign with the centre points; and for the teleportation spectator they show curvature of the opposite sign.

The second is the duration control, which measures the cost of the padding directly, by running the circuit with and without it. The trial run on \dev{ibm\_marrakesh} did this at one setting, and the follow-up at $\lL=1$ and 3, with four times as many shots. In the trial run, the $2t_{\rm slot}$ of padding lowered $\langle X_s\rangle$ by 0.064 (Sec.~\ref{sec:pilots}), that is, $\ln\langle X_s\rangle$ by 0.068 $[0.050,0.087]$, or 0.017 $[0.013,0.022]$ per microsecond. In the main experiment, the free evolution added by the latency knob cost only 0.0095 $[0.0089,0.0101]$ per microsecond ($-b_L/\tau_w$). The padding, which is part of the native window, thus decayed 1.8 $[1.3,2.4]$ times as fast as the stretch. In the follow-up it cost 0.055 $[0.049,0.061]$ at $\lL=1$, and decayed 1.5 $[1.3,1.7]$ times as fast as the stretch of the same run. If each slot of padding costs $p$ on the logarithmic scale and the stretch costs $r$ per unit time, every multi-knob estimate falls short of the value without a window by $3(p-rt_{\rm slot})$, plus a similar term for the feedforward wait, which we cannot vary (Appendix~\ref{app:derivations}). For the spectator, the first term alone is 0.046 $[0.019,0.075]$ from the trial run and 0.027 $[0.017,0.038]$ from the follow-up. On \dev{ibm\_kingston}, by contrast, the echoed padding lowered $\ln\langle X_s\rangle$ by only 0.007 $[-0.007,0.022]$, close to the 0.009 that the rate of the stretch there predicts. And for the probe's feedforward observable, the follow-up found no excess at all: its padding decayed 0.97 $[0.83,1.11]$ times as fast as its stretch, against 1.2 to 1.7 times in the trial run, depending on which run supplies the rate of the stretch.

For the spectator, the duration control directly establishes a higher average cost per unit time for the tested padding than for the stretch. The latency scan independently shows a nonconstant decay rate over the sampled range. Neither measurement separately identifies the unvaried feedforward wait, and interpreting them as a native-window mismatch requires the response assumptions stated above. For the feedforward observable the evidence is mixed. They are two estimates of the same shortfall, not corrections to be added. Under those models, the point estimates imply log shifts of 0.020 to 0.046, compared with an observed log deficit of 0.067 to 0.084. These are model-dependent scales, not an identified causal fraction of the total bias: the estimates span different runs and the wait contribution is unmeasured. The two more precise point estimates are 0.020 from the main latency scan and 0.027 from the follow-up duration control.

Why does the padding decay so fast? The simplest explanation would be that the first microseconds of any echoed delay decay fastest. That would also create an interaction. If only the total length of echoed delay mattered, the order within it would not: with one measurement the stretch adds to the $2t_{\rm slot}$ of padding, whereas with three it starts from no delay at all, and it would then cost more. In that case $b_{LM}=(c_3-c_1)/4$, where $c_1$ and $c_3$ are the costs of the padding at $\lL=1$ and 3 (Appendix~\ref{app:derivations}); and if, in addition, the decay is fastest at first, $c_3$ can be no larger than the cost $2rt_{\rm slot}$ of the same length of stretch, so that $b_{LM}\le-(p-rt_{\rm slot})/2$. Before the follow-up we could test only this bound, taking $p$ from the trial run and $r$ from the main experiment. For the spectator it is $-0.008$ $[-0.013,-0.003]$, whereas the main experiment found $+0.0007$, which seemed to rule the explanation out. The follow-up measured both costs in a single run. At $\lL=3$ the padding cost 0.042 $[0.031,0.054]$: less than at $\lL=1$, but a little more than the 0.037 that the rate of the stretch implies. The total-length prediction, $-0.0031$ $[-0.0063,+0.0002]$, agrees with the $-0.0014$ measured in the same run (difference $+0.0017$ $[-0.0017,+0.0052]$), and so does that for the feedforward observable ($-0.0026$ $[-0.0083,+0.0030]$, against $-0.0054$). But the spectator's measured value exceeds the bound for a decay that is fastest at first, $-0.0045$, by 0.0031 $[0.0003,0.0058]$, or $[-0.0007,+0.0070]$ with the bootstrap of our original plan. The follow-up thus shrinks the earlier contradiction, from 0.008 to 0.003, but does not clearly remove it, and the difference between the two costs, 0.012 $[-0.001,+0.025]$, is too uncertain to settle the matter: why the padding decays so fast remains open. If decay did depend only on the total length of delay, duration matching itself would contribute about $-0.003$ to the measured $b_{LM}$ of both probe observables. That would cut both ways. It would make the feedforward observable's $-0.0056$ in the main run, together with the spread between runs (Sec.~\ref{sec:repeat}), weak evidence that latency and measurement noise interact physically; and it would put the spectator's interaction in the main run, net of this artifact, at $+0.0038$ $[-0.0024,+0.0103]$ with the bootstrap of our original plan, or $[-0.0006,+0.0085]$ with the instance-level one. The bound of Sec.~\ref{sec:separability} applies to the knobs as built, duration matching included.

\subsection{Precision and an additive full-cube fit}
\label{sec:estimatorcomparison}
\begin{figure*}[t]
\centering
\includegraphics[width=0.94\textwidth]{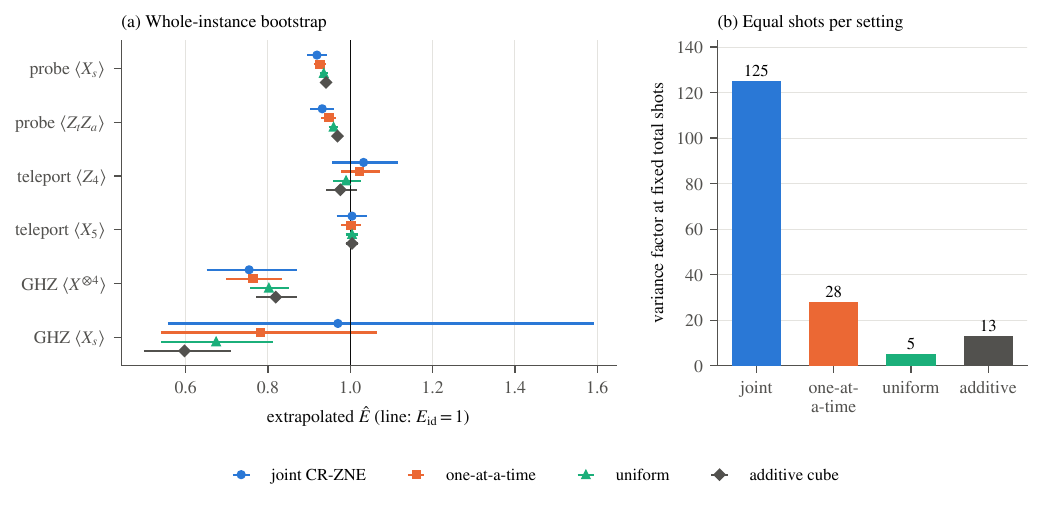}
\caption{Post-hoc precision analysis. (a) Extrapolated expectations from the main experiment; bars are 95\% whole-instance percentile bootstrap intervals with 20,000 replicates. The vertical line is the known ideal value, 1. These intervals describe within-run resampling uncertainty, not uncertainty from drift or response-model choice. (b) Analytic variance factors at a fixed total shot budget, with equal shots per required setting and a common per-shot log-variance factor. Panel (b) is a design calculation, not a rescaling of the intervals in (a).}
\label{fig:precision}
\end{figure*}

For a post-hoc comparison of estimator precision, we resampled the 20 complete twirled instances within each setting, independently across settings, using 20,000 replicates and seed 20261012. All 20,000 replicates were retained for each observable under the $E>0.05$ guard. This post-hoc analysis assumes exchangeable instances within a setting and does not model unobserved temporal correlation. It preserves the original outcomes and does not replace the pre-specified intervals. Figure~\ref{fig:precision} shows the resulting estimator intervals and the separate analytic shot-cost comparison.

The additive cube fit gives $0.941$ $[0.931,0.951]$ for the probe spectator, $0.969$ $[0.958,0.980]$ for probe feedforward, $0.976$ $[0.942,1.016]$ for teleportation, $1.004$ $[0.989,1.019]$ for its spectator, and $0.819$ $[0.771,0.870]$ for GHZ parity. The GHZ spectator is a counterexample to adopting it indiscriminately: its additive estimate is $0.597$ $[0.498,0.711]$, whereas the original joint point estimate is $0.970$. Its triple contrast, $b_{GLM}=-0.0473$, contributes strongly through Eq.~\eqref{eq:additivebias}. Fitting only main effects therefore trades variance for potentially large misspecification error. Choosing an estimator after looking at agreement with the known answer would require independent validation; these six outcomes do not establish an estimator-selection rule.

\section{Discussion}
\label{sec:discussion}

What have we learned? Within the resolution of the experiment, and once the idle periods are echoed, the primary interaction is consistent with zero: with the knobs as built, the latency\,$\times$\,measurement interaction seen by the spectator is smaller than 0.005 on the logarithmic scale. Elsewhere a few small interactions may be present. Two of the 24 interaction coefficients exclude zero in the pre-specified analysis, and four or five with the instance-level bootstrap of Sec.~\ref{sec:bootstrap}, against 1.2 expected by chance; except for the poorly determined GHZ spectator, every interaction is smaller in magnitude than the latency main effect. Accordingly, the simple strategies, one knob at a time or all knobs together, do about as well as the full joint fit. For these circuits, this agrees with the argument of Ref.~\cite{Etxezarreta2024} that a single global parameter can suffice, provided that every kind of noise that matters is amplified. In the teleportation and GHZ circuits, amplifying latency or gates alone left a much larger bias, which illustrates the risk of leaving a source of noise unamplified; in Ref.~\cite{Farrell2026}, for example, ZNE amplified gates and idling but not the mid-circuit resets. Our single-knob strategies leave gate or latency noise unamplified as well. In the probe, whose measured gate-knob effect is small, latency-only extrapolation did as well as any. Note also that uniform scaling needs the same three knobs as component-resolved scaling; what it saves is circuit settings, not instrumentation.

The clearest lesson may be methodological. Without the echo, our first trial run showed a log-scale interaction of 0.02, which could easily have been read as a physical effect; on the same device, the echo suppressed the rotation and reduced the interaction to a value consistent with zero. Carrera Vazquez \textit{et al.} had already seen that coherent oscillations spoil extrapolation in the latency~\cite{CarreraVazquez2024}; what Eq.~\eqref{eq:coh} adds is that they also masquerade as interactions. Indeed the effect is generic: whenever two knobs both shift a coherent phase, the logarithm of the observable acquires a cross term. Studies that look for interactions between amplified noise channels should therefore refocus static phases, or fit a model that allows for them, before interpreting the interactions. (Because the trial runs shaped our design, this comparison is exploratory.) The echo refocuses only the window, however. Phases picked up during the measurement slots do not depend on $\lL$, so by Eq.~\eqref{eq:coh} they can contribute to $b_{GM}$, if gate folding also shifts the phase, but not to $b_{LM}$. Whether this explains the one $b_{GM}$ whose interval excludes zero, for the teleportation spectator, we have not tested. The other interval that excludes zero, $b_{LM}$ for the probe's feedforward observable, did not recur consistently in later runs, and part of it may come from the design of the knobs: if decay depends only on the total length of echoed delay, which the follow-up neither confirms nor excludes, duration matching alone produces an interaction of about $-0.003$ (Secs.~\ref{sec:repeat} and~\ref{sec:window}). The coherent mechanism also supplies a cheap positive control. A virtual rotation applied with both signs injects an interaction of exactly known size, whatever phase the qubit already carries; our analysis recovered it at the size injected, at the cost of about one more set of the eight corners. But the same run carried a warning: two sets of circuits that should be equivalent for the feedforward observable gave values of its $b_{LM}$ that differ by 0.007, a difference that a permutation test made after the run puts at $p=0.02$ (Sec.~\ref{sec:separability}).

The measurement knob also departed from the textbook picture. Three repetitions raised the parity error by a factor of about 1.5 rather than 3, and later repetitions erred more often than earlier ones. Extrapolation along $\lM$ assumes that each repetition adds the same amount of noise; on this hardware it apparently does not, so the effective amplification factor differs from the nominal one.

Why, then, do the multi-knob estimates still fall short for half of the observables? We can offer three candidate explanations. The simplest is the gap between $E_*$ and $E_{\rm id}$: noise that no knob amplifies, such as that of single-qubit gates, state preparation, terminal readout and the echo pulses themselves. Readout alone is too small to explain the probe's deficit, however; the reported readout errors of its qubits, 0.0020 to 0.0046, would lower $\langle X_s\rangle$ by only about 0.004 and $\langle Z_tZ_a\rangle$ by about 0.018. A second candidate, and for the probe spectator the best supported, is that the latency knob does not stretch the window faithfully. As Sec.~\ref{sec:window} showed, the tested spectator padding has a higher average decay cost than the added stretch, and the latency scan is curved, with model-dependent log shifts comparable to part of the observed deficit. This does not identify a causal share of the shortfall. A nonconstant decay rate is one possible explanation, but the specific fastest-at-first model is in mild tension with one of the post-hoc intervals and is not established. For the GHZ parity, though, the latency scan shows no curvature, the two centre points disagree in sign, and we have no duration control for that circuit; curvature along $\lG$ and $\lM$ we could not measure at all. A third candidate is the measurement knob itself. If it amplifies measurement errors by less than its nominal factor on \dev{ibm\_marrakesh} too (we measured this only on \dev{ibm\_kingston}), extrapolation along $\lM$ removes only part of the feedforward error, which biases low the observables that depend on corrections but not the spectators; the teleportation observable, however, shows no such bias. Finally, the retargeting sensitivity analysis within the assumed range does not remove the shortfalls: changing it from 1 to 0.5 or 2~$\mu$s moves the probe's multi-knob estimates by less than 0.01 and the GHZ parity's by less than 0.025 (Sec.~\ref{sec:extrap}).

Agreement with the ideal value is not proof of correctness either. The teleportation spectator sits on qubit~9, whose reported readout error is 0.0156. No knob amplifies readout error, and because our terminal measurements are twirled, it acts as a symmetric bit flip, so by itself it should cap the extrapolated $\langle X\rangle$ at about $1-2\times0.0156\approx0.969$. Yet the multi-knob estimates are 1.002 to 1.004, above the cap: clearly so for uniform scaling, only just for one-at-a-time, and within the uncertainty for the joint fit. The positive curvature of this observable at both centre points would push the estimates upward, so the extrapolations may over-correct here; but the latency scan shows curvature of the opposite sign (Sec.~\ref{sec:window}), and the cap rests on a single calibration value, so the evidence is weak.

Our study has several limitations. The analysis was not registered externally, and the trial runs shaped the design. The main experiment ran on a different device from two of the trial runs. The circuits are small, with at most eight qubits, and come from a single family of devices. The measurement knob has only two levels and the gate knob's centre level is a mixture, so curvature along either cannot be seen. The extrapolations to $\lL=0$ depend on the assumed feedforward latency, which we varied only over the range our plan specified, 0.5 to 2~$\mu$s. The padding needed for duration matching raises the baseline idle noise, and on \dev{ibm\_marrakesh} it decays faster than the stretch, which our extrapolations do not correct; the two runs give model-dependent contributions of 0.046 and 0.027 to the log deficit. And folding lengthens the idle time before the measurement, which we did not compensate. Both repeat runs were flagged for drift, and the reported calibration values may not describe the device at the time of the runs (Appendix~\ref{app:runs}). With two levels per knob, our interactions describe non-additivity between the levels tested, not the whole response surface. Our decision rule for the primary interaction in the main run compared its uncertainty with the minimum detectable effect rather than testing equivalence; the test of equivalence in the follow-up used a margin fixed in advance, but chosen after the main run had produced it. The positive control injects a coherent interaction after the window; it shows that the analysis recovers an interaction of known size on this hardware, not that every physical mechanism of that size would appear in the same way. Two sets of circuits that should be equivalent for one observable gave values of its $b_{LM}$ that differ by 0.007 (Sec.~\ref{sec:separability}). And our bound on $b_{LM}$ applies to the knobs as built: if echoed delay decays according to its total length, part of any $b_{LM}$ measured with duration matching is an artifact of the design (Sec.~\ref{sec:window}). Above all, a small primary coefficient supports only a small value of that contrast at the tested settings. Observable separability requires all relevant mixed contrasts to be small; accurate mitigation additionally requires response and target assumptions.

\section{Conclusion}
\label{sec:conclusion}
A factorial experiment can diagnose the observable response to separately implemented noise-amplification knobs, but it cannot by itself certify a zero-noise estimate. The primary gate-averaged latency--measurement contrast is small on the echoed probe, and the planned follow-up meets its stated equivalence criterion while recovering injected interactions. The result concerns the knobs as implemented, including duration matching. It does not establish independence of physical channels, equivalence of every conditional contrast, or stability between calibrations.

The distinction matters experimentally: all five original estimators miss the known ideal value for three of six observables. Padding controls and latency scans reveal departures from a faithful scaling coordinate, while the two-level measurement design leaves its zero coordinate unidentifiable without further assumptions. A full joint fit pays a substantial variance cost, and the additive full-cube fit illustrates how reducing that cost can expose model bias. Neither complexity nor a small interaction coefficient is an accuracy guarantee.

A stronger next experiment would use additional physically realized amplification levels, an independently calibrated latency origin, contemporaneous padding controls for each circuit, and repeated randomized blocks spanning calibration changes. It should pre-specify an error tolerance for the total estimator-relevant contrast and allocate shots according to the chosen estimator's variance, before measuring ideal-value agreement on withheld circuits. The experiment establishes a reproducible diagnostic case study and identifies what those further measurements must resolve.

\section*{Data and code availability}
The ancillary files contain circuit and analysis code, raw counts, calibration snapshots, and code and inputs for the post-hoc analyses.

\section*{Use of AI tools}
Under the author's direction, Claude (Anthropic) assisted with editing, independent numerical checks, post-hoc analysis, and paraphrasing. The author takes responsibility for scientific content.

\begin{acknowledgments}
We acknowledge the use of IBM Quantum services for this work. The views expressed are those of the author, and do not reflect the official policy or position of IBM or the IBM Quantum team.
\end{acknowledgments}

\appendix

\section{Derivations}
\label{app:derivations}

Here we derive Eqs.~\eqref{eq:oat}--\eqref{eq:curv} from the model of Eq.~\eqref{eq:model}, and the results about the padding used in Sec.~\ref{sec:window}.

\paragraph*{Coded coefficients.} Inserting $\lambda_i=2+x_i$ into Eq.~\eqref{eq:model} and collecting terms, the coefficient of $x_ix_j$ ($i\neq j$) is $-\gamma_{ij}-2\gamma_{GLM}$; the factor 2 is the constant part of the third knob, $\lambda_k=2+x_k$, in the triple term. The coefficient of $x_Gx_Lx_M$ is $-\gamma_{GLM}$.

\paragraph*{One-at-a-time.} At $\mathbf e_i$ only $\lambda_i$ changes from 1 to 3. The linear term changes by $-2\gamma_i$, each pair term containing $i$ by $-2\gamma_{ij}$, and the triple term by $-2\gamma_{GLM}$. Hence $\hat\gamma_i=\gamma_i+\sum_{j\neq i}\gamma_{ij}+\gamma_{GLM}$. Summing,
\[
\hat f_0^{\rm OAT}=f(1,1,1)+\sum_i\hat\gamma_i=f_0+\sum_{i<j}\gamma_{ij}+2\gamma_{GLM},
\]
because $f(1,1,1)=f_0-\sum_i\gamma_i-\sum_{i<j}\gamma_{ij}-\gamma_{GLM}$ and each pair appears twice in $\sum_i\sum_{j\neq i}\gamma_{ij}$. Substituting $\gamma_{ij}=-b_{ij}+2b_{GLM}$ and $\gamma_{GLM}=-b_{GLM}$ gives Eq.~\eqref{eq:oat}.

\paragraph*{Uniform.} Along the diagonal, $f(\lambda)=f_0-\Gamma\lambda-S\lambda^2-\gamma_{GLM}\lambda^3$ with $\Gamma=\sum_i\gamma_i$ and $S=\sum_{i<j}\gamma_{ij}$. The two-point slope is $[f(3)-f(1)]/2=-\Gamma-4S-13\gamma_{GLM}$, so
\[
\hat f_0^{\rm unif}=f(1)-[f(3)-f(1)]/2=f_0+3S+12\gamma_{GLM},
\]
which is Eq.~\eqref{eq:unif} after the same substitution.

\paragraph*{Coherent phase.} With $\varphi=\varphi_0+\sum_ia_ix_i$, $-\varphi^2/2$ contains $-\tfrac12\cdot2a_La_Mx_Lx_M$. Higher orders of $\ln\cos\varphi$ add terms of order $\varphi^4$, so Eq.~\eqref{eq:coh} is the leading-order contribution.

\paragraph*{Curvature.} Add pure quadratic terms to the model, $f\to f+\sum_ik_i\lambda_i^2$. On two levels, $k_i\lambda_i^2$ takes the values $k_i$ and $9k_i$ at $\lambda_i=1$ and 3. Every multi-knob estimator extrapolates it linearly to $\lambda_i=0$, which gives $k_i-4k_i=-3k_i$ instead of 0, so each estimate of $f_0$ is shifted by $-3\sum_ik_i$. At the centre point $(2,2,1)$ the quadratic terms contribute $4k_G+4k_L+k_M$; the mean of the four $\lM=1$ corners contributes $5k_G+5k_L+k_M$, while linear and interaction terms contribute equally to both. The curvature is therefore $c=-(k_G+k_L)$, and the shift is $3c-3k_M$.

\paragraph*{Mixed centre point.} If $\lG=2$ is realized as an equal mixture of $\lG=1$ and $\lG=3$ circuits, as in our experiment, then to first order the centre point sees the average of $k_G$ and $9k_G$, that is $5k_G$, just as the corners do. At this order the curvature is $c\simeq-k_L$, and the inferred shift is approximately $3c-3k_G-3k_M$. In the experiment the mixtures were not always equal: by chance, 7 of the 20 instances of the probe's $(2,2,1)$ circuit were folded, and 8 of 20 for the teleportation circuit at $(2,2,3)$. The linear gate terms then shift $c$ by $b_G\,\delta x_G+b_{GM}\,\delta x_G\,x_M$, where $\delta x_G$ is the departure of the mean $\lG$ from 2; for our observables this is at most about 0.004, well within the uncertainties. Beyond first order, the logarithm of an average exceeds the average of the logarithms: for an equal mixture, $f_{\rm mix}=(f_1+f_3)/2+\ln\cosh(\Delta/2)$, where $f_1$ and $f_3$ are the values of $f$ for the two components and $\Delta=f_3-f_1=2(b_G+b_{GM}x_M)$ at the centre. This raises $c$ by about $(b_G+b_{GM}x_M)^2/2$, at most 0.002 for our observables (the GHZ parity at $(2,2,3)$), again well within the uncertainties.

\paragraph*{Padding.} Let one slot of padding cost $p$ on the logarithmic scale, a measurement slot $m$ and the feedforward wait $w$, and let the free evolution added by the latency knob cost $r$ per unit time. With duration matching, the window at $(\lL,\lM)$ holds $\lM$ measurement slots, $3-\lM$ slots of padding, the wait and $(\lL-1)\tau_w$ of added evolution, so that $f=f_{\rm nw}-m\lM-p(3-\lM)-w-r\tau_w(\lL-1)$, where $f_{\rm nw}$ is the value without a window. This is linear in each knob, so every multi-knob strategy extrapolates it exactly, to $f_{\rm nw}-3p-w+r\tau_w$. Because $\tau_w=3t_{\rm slot}+\tau_{\rm ff}$, the estimate falls short of $f_{\rm nw}$ by $3(p-rt_{\rm slot})+(w-r\tau_{\rm ff})$, which vanishes if the padding and the wait decay at the rate of the stretch; the cost $m$ of the measurements drops out. The duration control measures $2p$. If instead $f$ depended only on the total length $D=(3-\lM)t_{\rm slot}+(\lL-1)\tau_w$ of echoed free evolution in the window, the four corners at either $\lG$ would give $4b_{LM}=[f(2t_{\rm slot})-f(0)]-[f(2t_{\rm slot}+2\tau_w)-f(2\tau_w)]$, the change caused by the padding at the start of the delay minus the change caused by the same length of delay after $2\tau_w$ of it. Taking the first from the duration control, $-2p$, and the second at the rate of the stretch, $-2rt_{\rm slot}$, gives $b_{LM}=-(p-rt_{\rm slot})/2$. If the decay is fastest at first, the cost of the later delay can be no larger than $2rt_{\rm slot}$, and this value is an upper bound on $b_{LM}$. The follow-up measured both changes directly, as the costs $c_1$ and $c_3$ of the padding at $\lL=1$ and 3, which gives $b_{LM}=(c_3-c_1)/4$.

\section{Device calibration}
\label{app:runs}

The text quotes a few values from the calibration snapshots recorded at submission. For the main run, the reported terminal-readout errors were 0.0020 to 0.0046 on the probe's qubits (4--6 on \dev{ibm\_marrakesh}) and 0.0156 on qubit~9, which carries the teleportation spectator. For the repeat run, $T_1$ and $T_2$ changed, notably on qubit~8 ($T_1$ from 214 to 33~$\mu$s and $T_2$ from 148 to 59~$\mu$s), which the probe does not use; but the reported readout and gate errors were identical to those of the main run, so they may not have been refreshed. Before the follow-up, the reported readout errors of qubits 4 and 5 were 0.0081 and 0.0096, against 0.0046 and 0.0043 for the main run; the error of mid-circuit measurements on qubit 5 was 0.0186, against 0.0020; and the reported CZ errors had fallen by about a third. The readout error reported for qubit~11 (0.2708) is inconsistent with our data: the GHZ spectator measured on that qubit gave $\langle X\rangle=0.6215$ in the native circuit, whereas a twirled readout error of 0.2708 alone would cap it at $1-2\times0.2708\approx0.46$. We do not use that value. The full snapshots are part of the data set.

\end{document}